%% file: weather_sensing_paper.tex
\documentclass[conference]{IEEEtran}
\IEEEoverridecommandlockouts
\usepackage[letterpaper, left = 0.68in, right =0.68in, top = 0.7in, bottom = 1in]{geometry}
\usepackage{cite}
\usepackage[T1]{fontenc}
\usepackage{graphicx}
\usepackage{amssymb}
\usepackage{amsmath}
\usepackage{amsthm}
\usepackage{booktabs}
\usepackage{multirow} 
\usepackage{microtype}
\usepackage{balance}
\usepackage{xcolor}
\usepackage[acronym]{glossaries}
\usepackage{tikz}
\usetikzlibrary{arrows.meta, positioning}
\usepackage{subcaption}

\input{sections/acronyms}

\makeglossaries

\input{macros/vmr-symbols-vecbold}
\input{macros/standard-macros}

\input{macros/defs}

\newcommand*{\fancyrefremarklabelprefix}{remark}
\frefformat{vario}{\fancyrefremarklabelprefix}{Remark~#1}

\newcommand{\PR}[1]{\ensuremath{\!\left[#1\right]}}
\newcommand{\PC}[1]{\ensuremath{\!\left(#1\right)}}
\newcommand{\chav}[1]{\ensuremath{\!\left\{#1\right\}}}
\newcommand{\PM}[1]{\ensuremath{\!\left|#1\right|}}

\usepackage{color}

\def\loss{\mathfrak{L}}

\IEEEoverridecommandlockouts 
\allowdisplaybreaks 


\title{Weather Estimation for Integrated Sensing and Communication
\author{\IEEEauthorblockN{Victoria Palhares, Artjom Grudnitsky, and Silvio Mandelli}
\IEEEauthorblockA{Nokia Bell Labs Stuttgart, Germany\\
Email: \{victoria.palhares, artjom.grudnitsky, silvio.mandelli\}@nokia-bell-labs.com}}}

\begin{document}

\maketitle

\input{sections/0-abstract}

\input{sections/1-introduction}
\input{sections/2-periodogram_generation}
\input{sections/3-weather_sensing}

\input{sections/4-results}

\input{sections/5-conclusions}

\bibliographystyle{IEEEtran}

\balance

\bibliography{bib/IEEEabrv,bib/confs-jrnls,bib/publishers,bib/weather_sensing} 

\balance 
	
\end{document}

%% file: sections/acronyms.tex
\newacronym{5g}{5G}{fifth-generation}
\newacronym{6g}{6G}{sixth-generation}
\newacronym[plural=BSs,firstplural = base stations (BSs)]{bs}{BS}{base station}
\newacronym[plural=CDFs,firstplural = cumulative distribution function (CDFs)]{cdf}{CDF}{cumulative distribution function}
\newacronym[plural=CNNs,firstplural = convolutional neural networks (CNNs)]{cnn}{CNN}{convolutional neural network}
\newacronym[plural=ConvLSTMs,firstplural = convolutional long-short term memory networks (ConvLSTMs)]{convlstm}{ConvLSTM}{convolutional long-short term memory network}
\newacronym{crap}{CRAP}{clutter removal with acquisitions under phase noise}
\newacronym{csi}{CSI}{channel state information}
\newacronym{eirp}{EIRP}{effective isotropic radiated power}
\newacronym{em}{EM}{electromagnetic}
\newacronym[plural=FFTs,firstplural = fast Fourier transform (FFTs)]{fft}{FFT}{fast Fourier transform}
\newacronym[plural=gNB-RUs,firstplural = next-generation node B-radio units (gNB-RUs)]{gnbru}{gNB-RU}{next-generation node B-radio unit}
\newacronym[plural=GRUs,firstplural = gated recurrent units (GRUs)]{gru}{GRU}{gated recurrent unit}
\newacronym[plural=IFFTs,firstplural = inverse FFTs (IFFTs)]{ifft}{IFFT}{inverse FFT}
\newacronym{ecpri}{eCPRI}{evolved common public radio interface}
\newacronym{inp}{InP}{interferometric phase}
\newacronym{isac}{ISAC}{integrated sensing and communication}
\newacronym{lidar}{LIDAR}{light detection and recognition}
\newacronym{los}{LoS}{line-of-sight}
\newacronym[plural=LSTMs,longplural = long short-term memory networks]{lstm}{LSTM}{long short-term memory}
\newacronym{mae}{MAE}{mean absolute error}
\newacronym{mse}{MSE}{mean squared error}
\newacronym{mmwave}{mmWave}{millimeter}
\newacronym[plural=NNs,firstplural = neural networks (NNs)]{nn}{NN}{neural network}
\newacronym{ofdm}{OFDM}{orthogonal frequency-division multiplexing}
\newacronym{1d}{1D}{one-dimensional}
\newacronym{poc}{PoC}{proof-of-concept}
\newacronym[plural=RMSE,firstplural = root mean squared errors (RMSEs)]{rmse}{RMSE}{root mean squared error}
\newacronym[plural=RNNs,firstplural = recurrent neural networks (RNNs)]{rnn}{RNN}{recurrent neural network}
\newacronym[plural=RUs,firstplural = radio units (RUs)]{ru}{RU}{radio unit}
\newacronym[plural=Rxs,firstplural = receivers (Rxs)]{rx}{Rx}{receiver}
\newacronym[plural=SLPs,firstplural = single-layer perceptrons (SLPs)]{slp}{SLP}{single-layer perceptron}
\newacronym[plural=SVMs,firstplural = support vector machines (SVMs)]{svm}{SVM}{support vector machine}
\newacronym[plural=Txs,firstplural = transmitters (Txs)]{tx}{Tx}{transmitter}
\newacronym{spu}{SPU}{sensing processing unit}
\newacronym{2d}{2D}{two-dimensional}
\newacronym{3d}{3D}{three-dimensional}
\newacronym[plural=UEs,firstplural = user equipments (UEs)]{ue}{UE}{user equipment}
\newacronym[plural=URAs,firstplural = uniform rectangular arrays (URAs)]{ura}{URA}{uniform rectangular array}

%% file: macros/vmr-symbols-vecbold.tex
\usepackage{amssymb}
\usepackage{amsfonts}
\usepackage{mathrsfs}
\usepackage{xspace}
\usepackage{bm}
\usepackage{upgreek}

\newcommand{\safemath}[2]{\newcommand{#1}{\ensuremath{#2}\xspace}}

\safemath{\bma}{\mathbf{a}}
\safemath{\bmb}{\mathbf{b}}
\safemath{\bmc}{\mathbf{c}}
\safemath{\bmd}{\mathbf{d}}
\safemath{\bme}{\mathbf{e}}
\safemath{\bmf}{\mathbf{f}}
\safemath{\bmg}{\mathbf{g}}
\safemath{\bmh}{\mathbf{h}}
\safemath{\bmi}{\mathbf{i}}
\safemath{\bmj}{\mathbf{j}}
\safemath{\bmk}{\mathbf{k}}
\safemath{\bml}{\mathbf{l}}
\safemath{\bmm}{\mathbf{m}}
\safemath{\bmn}{\mathbf{n}}
\safemath{\bmo}{\mathbf{o}}
\safemath{\bmp}{\mathbf{p}}
\safemath{\bmq}{\mathbf{q}}
\safemath{\bmr}{\mathbf{r}}
\safemath{\bms}{\mathbf{s}}
\safemath{\bmt}{\mathbf{t}}
\safemath{\bmu}{\mathbf{u}}
\safemath{\bmv}{\mathbf{v}}
\safemath{\bmw}{\mathbf{w}}
\safemath{\bmx}{\mathbf{x}}
\safemath{\bmy}{\mathbf{y}}
\safemath{\bmz}{\mathbf{z}}
\safemath{\bmzero}{\mathbf{0}}
\safemath{\bmone}{\mathbf{1}}

\bmdefine{\biad}{a}
\bmdefine{\bibd}{b}
\bmdefine{\bicd}{c}
\bmdefine{\bidd}{d}
\bmdefine{\bied}{e}
\bmdefine{\bifd}{f}
\bmdefine{\bigd}{g}
\bmdefine{\bihd}{h}
\bmdefine{\biid}{i}
\bmdefine{\bijd}{j}
\bmdefine{\bikd}{k}
\bmdefine{\bild}{l}
\bmdefine{\bimd}{m}
\bmdefine{\bind}{n}
\bmdefine{\biod}{o}
\bmdefine{\bipd}{p}
\bmdefine{\biqd}{q}
\bmdefine{\bird}{r}
\bmdefine{\bisd}{s}
\bmdefine{\bitd}{t}
\bmdefine{\biud}{u}
\bmdefine{\bivd}{v}
\bmdefine{\biwd}{w}
\bmdefine{\bixd}{x}
\bmdefine{\biyd}{y}
\bmdefine{\bizd}{z}

\bmdefine{\bixid}{\xi}
\bmdefine{\bilambdad}{\lambda}
\bmdefine{\bimud}{\mu}
\bmdefine{\bithetad}{\theta}
\bmdefine{\biphid}{\phi}
\bmdefine{\bideltad}{\delta}

\safemath{\bmia}{\biad}
\safemath{\bmib}{\bibd}
\safemath{\bmic}{\bicd}
\safemath{\bmid}{\bidd}
\safemath{\bmie}{\bied}
\safemath{\bmif}{\bifd}
\safemath{\bmig}{\bigd}
\safemath{\bmih}{\bihd}
\safemath{\bmii}{\biid}
\safemath{\bmij}{\bijd}
\safemath{\bmik}{\bikd}
\safemath{\bmil}{\bild}
\safemath{\bmim}{\bimd}
\safemath{\bmin}{\bind}
\safemath{\bmio}{\biod}
\safemath{\bmip}{\bipd}
\safemath{\bmiq}{\biqd}
\safemath{\bmir}{\bird}
\safemath{\bmis}{\bisd}
\safemath{\bmit}{\bitd}
\safemath{\bmiu}{\biud}
\safemath{\bmiv}{\bivd}
\safemath{\bmiw}{\biwd}
\safemath{\bmix}{\bixd}
\safemath{\bmiy}{\biyd}
\safemath{\bmiz}{\bizd}

\safemath{\bmxi}{\bixid}
\safemath{\bmlambda}{\bilambdad}
\safemath{\bmmu}{\bimud}
\safemath{\bmtheta}{\bithetad}
\safemath{\bmphi}{\biphid}
\safemath{\bmdelta}{\bideltad}

\safemath{\bA}{\mathbf{A}}
\safemath{\bB}{\mathbf{B}}
\safemath{\bC}{\mathbf{C}}
\safemath{\bD}{\mathbf{D}}
\safemath{\bE}{\mathbf{E}}
\safemath{\bF}{\mathbf{F}}
\safemath{\bG}{\mathbf{G}}
\safemath{\bH}{\mathbf{H}}
\safemath{\bI}{\mathbf{I}}
\safemath{\bJ}{\mathbf{J}}
\safemath{\bK}{\mathbf{K}}
\safemath{\bL}{\mathbf{L}}
\safemath{\bM}{\mathbf{M}}
\safemath{\bN}{\mathbf{N}}
\safemath{\bO}{\mathbf{O}}
\safemath{\bP}{\mathbf{P}}
\safemath{\bQ}{\mathbf{Q}}
\safemath{\bR}{\mathbf{R}}
\safemath{\bS}{\mathbf{S}}
\safemath{\bT}{\mathbf{T}}
\safemath{\bU}{\mathbf{U}}
\safemath{\bV}{\mathbf{V}}
\safemath{\bW}{\mathbf{W}}
\safemath{\bX}{\mathbf{X}}
\safemath{\bY}{\mathbf{Y}}
\safemath{\bZ}{\mathbf{Z}}

\safemath{\bZero}{\mathbf{0}}
\safemath{\bOne}{\mathbf{1}}
\safemath{\bDelta}{\mathbf{\Delta}}
\safemath{\bLambda}{\mathbf{\UpLambda}}
\safemath{\bPhi}{\mathbf{\Upphi}}
\safemath{\bSigma}{\mathbf{\Upsigma}}
\safemath{\bOmega}{\mathbf{\Upomega}}
\safemath{\bTheta}{\mathbf{\Uptheta}}

\bmdefine{\biAd}{A}
\bmdefine{\biBd}{B}
\bmdefine{\biCd}{C}
\bmdefine{\biDd}{D}
\bmdefine{\biEd}{E}
\bmdefine{\biFd}{F}
\bmdefine{\biGd}{G}
\bmdefine{\biHd}{H}
\bmdefine{\biId}{I}
\bmdefine{\biJd}{J}
\bmdefine{\biKd}{K}
\bmdefine{\biLd}{L}
\bmdefine{\biMd}{M}
\bmdefine{\biOd}{N}
\bmdefine{\biPd}{O}
\bmdefine{\biQd}{P}
\bmdefine{\biRd}{R}
\bmdefine{\biSd}{S}
\bmdefine{\biTd}{T}
\bmdefine{\biUd}{U}
\bmdefine{\biVd}{V}
\bmdefine{\biWd}{W}
\bmdefine{\biXd}{X}
\bmdefine{\biYd}{Y}
\bmdefine{\biZd}{Z}

\bmdefine{\biDelta}{\Delta}
\bmdefine{\biLambda}{\Lambda}
\bmdefine{\biPhi}{\Phi}
\bmdefine{\biSigma}{\Sigma}
\bmdefine{\biOmega}{\Omega}
\bmdefine{\biTheta}{\Theta}

\safemath{\bimA}{\biAd}
\safemath{\bimB}{\biBd}
\safemath{\bimC}{\biCd}
\safemath{\bimD}{\biDd}
\safemath{\bimE}{\biEd}
\safemath{\bimF}{\biFd}
\safemath{\bimG}{\biGd}
\safemath{\bimH}{\biHd}
\safemath{\bimI}{\biId}
\safemath{\bimJ}{\biJd}
\safemath{\bimK}{\biKd}
\safemath{\bimL}{\biLd}
\safemath{\bimM}{\biMd}
\safemath{\bimN}{\biNd}
\safemath{\bimO}{\biOd}
\safemath{\bimP}{\biPd}
\safemath{\bimQ}{\biQd}
\safemath{\bimR}{\biRd}
\safemath{\bimS}{\biSd}
\safemath{\bimT}{\biTd}
\safemath{\bimU}{\biUd}
\safemath{\bimV}{\biVd}
\safemath{\bimW}{\biWd}
\safemath{\bimX}{\biXd}
\safemath{\bimY}{\biYd}
\safemath{\bimZ}{\biZd}

\safemath{\bimDelta}{\biDelta}
\safemath{\bimLambda}{\biLambda}
\safemath{\bimPhi}{\biPhi}
\safemath{\bimSigma}{\biSigma}
\safemath{\bimOmega}{\biOmega}
\safemath{\bimTheta}{\biTheta}

\safemath{\setA}{\mathcal{A}}
\safemath{\setB}{\mathcal{B}}
\safemath{\setC}{\mathcal{C}}
\safemath{\setD}{\mathcal{D}}
\safemath{\setE}{\mathcal{E}}
\safemath{\setF}{\mathcal{F}}
\safemath{\setG}{\mathcal{G}}
\safemath{\setH}{\mathcal{H}}
\safemath{\setI}{\mathcal{I}}
\safemath{\setJ}{\mathcal{J}}
\safemath{\setK}{\mathcal{K}}
\safemath{\setL}{\mathcal{L}}
\safemath{\setM}{\mathcal{M}}
\safemath{\setN}{\mathcal{N}}
\safemath{\setO}{\mathcal{O}}
\safemath{\setP}{\mathcal{P}}
\safemath{\setQ}{\mathcal{Q}}
\safemath{\setR}{\mathcal{R}}
\safemath{\setS}{\mathcal{S}}
\safemath{\setT}{\mathcal{T}}
\safemath{\setU}{\mathcal{U}}
\safemath{\setV}{\mathcal{V}}
\safemath{\setW}{\mathcal{W}}
\safemath{\setX}{\mathcal{X}}
\safemath{\setY}{\mathcal{Y}}
\safemath{\setZ}{\mathcal{Z}}
\safemath{\emptySet}{\varnothing}

\safemath{\colA}{\mathscr{A}}
\safemath{\colB}{\mathscr{B}}
\safemath{\colC}{\mathscr{C}}
\safemath{\colD}{\mathscr{D}}
\safemath{\colE}{\mathscr{E}}
\safemath{\colF}{\mathscr{F}}
\safemath{\colG}{\mathscr{G}}
\safemath{\colH}{\mathscr{H}}
\safemath{\colI}{\mathscr{I}}
\safemath{\colJ}{\mathscr{J}}
\safemath{\colK}{\mathscr{K}}
\safemath{\colL}{\mathscr{L}}
\safemath{\colM}{\mathscr{M}}
\safemath{\colN}{\mathscr{N}}
\safemath{\colO}{\mathscr{O}}
\safemath{\colP}{\mathscr{P}}
\safemath{\colQ}{\mathscr{Q}}
\safemath{\colR}{\mathscr{R}}
\safemath{\colS}{\mathscr{S}}
\safemath{\colT}{\mathscr{T}}
\safemath{\colU}{\mathscr{U}}
\safemath{\colV}{\mathscr{V}}
\safemath{\colW}{\mathscr{W}}
\safemath{\colX}{\mathscr{X}}
\safemath{\colY}{\mathscr{Y}}
\safemath{\colZ}{\mathscr{Z}}

\safemath{\opA}{\mathbb{A}}
\safemath{\opB}{\mathbb{B}}
\safemath{\opC}{\mathbb{C}}
\safemath{\opD}{\mathbb{D}}
\safemath{\opE}{\mathbb{E}}
\safemath{\opF}{\mathbb{F}}
\safemath{\opG}{\mathbb{G}}
\safemath{\opH}{\mathbb{H}}
\safemath{\opI}{\mathbb{I}}
\safemath{\opJ}{\mathbb{J}}
\safemath{\opK}{\mathbb{K}}
\safemath{\opL}{\mathbb{L}}
\safemath{\opM}{\mathbb{M}}
\safemath{\opN}{\mathbb{N}}
\safemath{\opO}{\mathbb{O}}
\safemath{\opP}{\mathbb{P}}
\safemath{\opQ}{\mathbb{Q}}
\safemath{\opR}{\mathbb{R}}
\safemath{\opS}{\mathbb{S}}
\safemath{\opT}{\mathbb{T}}
\safemath{\opU}{\mathbb{U}}
\safemath{\opV}{\mathbb{V}}
\safemath{\opW}{\mathbb{W}}
\safemath{\opX}{\mathbb{X}}
\safemath{\opY}{\mathbb{Y}}
\safemath{\opZ}{\mathbb{Z}}
\safemath{\opZero}{\mathbb{O}}
\safemath{\identityop}{\opI}

\safemath{\veca}{\bma}
\safemath{\vecb}{\bmb}
\safemath{\vecc}{\bmc}
\safemath{\vecd}{\bmd}
\safemath{\vece}{\bme}
\safemath{\vecf}{\bmf}
\safemath{\vecg}{\bmg}
\safemath{\vech}{\bmh}
\safemath{\veci}{\bmi}
\safemath{\vecj}{\bmj}
\safemath{\veck}{\bmk}
\safemath{\vecl}{\bml}
\safemath{\vecm}{\bmm}
\safemath{\vecn}{\bmn}
\safemath{\veco}{\bmo}
\safemath{\vecp}{\bmp}
\safemath{\vecq}{\bmq}
\safemath{\vecr}{\bmr}
\safemath{\vecs}{\bms}
\safemath{\vect}{\bmt}
\safemath{\vecu}{\bmu}
\safemath{\vecv}{\bmv}
\safemath{\vecw}{\bmw}
\safemath{\vecx}{\bmx}
\safemath{\vecy}{\bmy}
\safemath{\vecz}{\bmz}

\safemath{\veczero}{\bmzero}
\safemath{\vecone}{\bmone}
\safemath{\vecxi}{\bmxi}
\safemath{\veclambda}{\bmlambda}
\safemath{\vecmu}{\bmmu}
\safemath{\vectheta}{\bmtheta}
\safemath{\vecphi}{\bmphi}
\safemath{\vecdelta}{\bmdelta}

\safemath{\matA}{\bA}
\safemath{\matB}{\bB}
\safemath{\matC}{\bC}
\safemath{\matD}{\bD}
\safemath{\matE}{\bE}
\safemath{\matF}{\bF}
\safemath{\matG}{\bG}
\safemath{\matH}{\bH}
\safemath{\matI}{\bI}
\safemath{\matJ}{\bJ}
\safemath{\matK}{\bK}
\safemath{\matL}{\bL}
\safemath{\matM}{\bM}
\safemath{\matN}{\bN}
\safemath{\matO}{\bO}
\safemath{\matP}{\bP}
\safemath{\matQ}{\bQ}
\safemath{\matR}{\bR}
\safemath{\matS}{\bS}
\safemath{\matT}{\bT}
\safemath{\matU}{\bU}
\safemath{\matV}{\bV}
\safemath{\matW}{\bW}
\safemath{\matX}{\bX}
\safemath{\matY}{\bY}
\safemath{\matZ}{\bZ}
\safemath{\matzero}{\bmzero}

\safemath{\matDelta}{\bDelta}
\safemath{\matLambda}{\bLambda}
\safemath{\matPhi}{\bPhi}
\safemath{\matSigma}{\bSigma}
\safemath{\matOmega}{\bOmega}
\safemath{\matTheta}{\bTheta}

\safemath{\matidentity}{\matI}
\safemath{\matone}{\matO}

\safemath{\rnda}{A}
\safemath{\rndb}{B}
\safemath{\rndc}{C}
\safemath{\rndd}{D}
\safemath{\rnde}{E}
\safemath{\rndf}{F}
\safemath{\rndg}{G}
\safemath{\rndh}{H}
\safemath{\rndi}{I}
\safemath{\rndj}{J}
\safemath{\rndk}{K}
\safemath{\rndl}{L}
\safemath{\rndm}{M}
\safemath{\rndn}{N}
\safemath{\rndo}{O}
\safemath{\rndp}{P}
\safemath{\rndq}{Q}
\safemath{\rndr}{R}
\safemath{\rnds}{S}
\safemath{\rndt}{T}
\safemath{\rndu}{U}
\safemath{\rndv}{V}
\safemath{\rndw}{W}
\safemath{\rndx}{X}
\safemath{\rndy}{Y}
\safemath{\rndz}{Z}

\safemath{\rveca}{\bimA}
\safemath{\rvecb}{\bimB}
\safemath{\rvecc}{\bimC}
\safemath{\rvecd}{\bimD}
\safemath{\rvece}{\bimE}
\safemath{\rvecf}{\bimF}
\safemath{\rvecg}{\bimG}
\safemath{\rvech}{\bimH}
\safemath{\rveci}{\bimI}
\safemath{\rvecj}{\bimJ}
\safemath{\rveck}{\bimK}
\safemath{\rvecl}{\bimL}
\safemath{\rvecm}{\bimM}
\safemath{\rvecn}{\bimN}
\safemath{\rveco}{\bomO}
\safemath{\rvecp}{\bimP}
\safemath{\rvecq}{\bimQ}
\safemath{\rvecr}{\bimR}
\safemath{\rvecs}{\bimS}
\safemath{\rvect}{\bimT}
\safemath{\rvecu}{\bimU}
\safemath{\rvecv}{\bimV}
\safemath{\rvecw}{\bimW}
\safemath{\rvecx}{\bimX}
\safemath{\rvecy}{\bimY}
\safemath{\rvecz}{\bimZ}

\safemath{\rvecxi}{\bmxi}
\safemath{\rveclambda}{\bmlambda}
\safemath{\rvecmu}{\bmmu}
\safemath{\rvectheta}{\bmtheta}
\safemath{\rvecphi}{\bmphi}

\safemath{\rmatA}{\bimA}
\safemath{\rmatB}{\bimB}
\safemath{\rmatC}{\bimC}
\safemath{\rmatD}{\bimD}
\safemath{\rmatE}{\bimE}
\safemath{\rmatF}{\bimF}
\safemath{\rmatG}{\bimG}
\safemath{\rmatH}{\bimH}
\safemath{\rmatI}{\bimI}
\safemath{\rmatJ}{\bimJ}
\safemath{\rmatK}{\bimK}
\safemath{\rmatL}{\bimL}
\safemath{\rmatM}{\bimM}
\safemath{\rmatN}{\bimN}
\safemath{\rmatO}{\bimO}
\safemath{\rmatP}{\bimP}
\safemath{\rmatQ}{\bimQ}
\safemath{\rmatR}{\bimR}
\safemath{\rmatS}{\bimS}
\safemath{\rmatT}{\bimT}
\safemath{\rmatU}{\bimU}
\safemath{\rmatV}{\bimV}
\safemath{\rmatW}{\bimW}
\safemath{\rmatX}{\bimX}
\safemath{\rmatY}{\bimY}
\safemath{\rmatZ}{\bimZ}

\safemath{\rmatDelta}{\bimDelta}
\safemath{\rmatLambda}{\bimLambda}
\safemath{\rmatPhi}{\bimPhi}
\safemath{\rmatSigma}{\bimSigma}
\safemath{\rmatOmega}{\bimOmega}
\safemath{\rmatTheta}{\bimTheta}

%% file: macros/standard-macros.tex
\usepackage{amssymb}
\usepackage{amsfonts}
\usepackage{mathrsfs}
\usepackage{xspace}
\usepackage{bm}
\usepackage{fancyref}
\usepackage{textcomp}

\usepackage{multirow}
\usepackage{stmaryrd}

\newenvironment{textbmatrix}{	\setlength{\arraycolsep}{2.5pt}%
								\big[\begin{matrix}}{\end{matrix}\big]%
								\raisebox{0.08ex}{\vphantom{M}}}

\def\be{\begin{equation}}
\def\ee{\end{equation}}
\def\een{\nonumber \end{equation}}
\def\mat{\begin{bmatrix}}
\def\emat{\end{bmatrix}}
\def\btm{\begin{textbmatrix}}
\def\etm{\end{textbmatrix}}

\def\ba#1\ea{\begin{align}#1\end{align}}
\def\bas#1\eas{\begin{align*}#1\end{align*}}
\def\bs#1\es{\begin{split}#1\end{split}}
\def\bg#1\eg{\begin{gather}#1\end{gather}}
\def\bml#1\eml{\begin{multline}#1\end{multline}}
\def\bi#1\ei{\begin{itemize}#1\end{itemize}}

\DeclareMathOperator*{\argmax}{arg\;max}		% arg max
\safemath{\dirac}{\delta}					% Dirac delta
\safemath{\krond}{\dirac}					% Kronecker delta
\safemath{\upto}{\uparrow}
\safemath{\downto}{\downarrow}
\safemath{\iu}{j}							% imaginary unit
\safemath{\ev}{\lambda}						% eigenvalue
\safemath{\hilseqspace}{l^{2}}				% Hilbert sequence space
\newcommand{\banachfunspace}[1]{\setL^{#1}}	% Banach function space
\safemath{\hilfunspace}{\banachfunspace{2}}	% Hilbert function space
\newcommand{\floor}[1]{\lfloor #1 \rfloor}
\newcommand{\ceil}[1]{\lceil #1 \rceil}

\safemath{\SNR}{\textit{SNR}} 				% signal to noise ratio
\safemath{\PAR}{\textit{PAR}} 				% signal to noise ratio
\safemath{\No}{N_0}							% noise spectral density
\safemath{\Es}{E_s}							% energy per symbol
\safemath{\Eb}{E_b}							% energy per bit
\safemath{\EbNo}{\frac{\Eb}{\No}}
\safemath{\EsNo}{\frac{\Es}{\No}}

\DeclareMathOperator{\CHop}{\ensuremath{\opH}} % channel operator
\safemath{\tvir}{\rndh_{\CHop}}				% time-varying impulse response
\safemath{\tvtf}{\rndl_{\CHop}}				% 	-''- transfer function
\safemath{\spf}{\rnds_{\CHop}}				% spreading function
\safemath{\bff}{H_{\CHop}}					% bi-freuqency function

\safemath{\ircf}{r_{h}}						% impulse response correlation fn.
\safemath{\tftvcf}{r_{s}}					% scattering function
\safemath{\tfcf}{r_{l}}						% time-frequency correlation fn.
\safemath{\bfcf}{r_{H}}						% bi-frequency correlation fn.

\safemath{\tcorr}{c_h}						% time-correlation function
\safemath{\scf}{c_{s}}						% spreading function
\safemath{\tfcorr}{c_{l}}					% transfer-function correlation
\safemath{\fcorr}{c_{H}}						% frequency-correlation function

\safemath{\mi}{I}							% mutual information
\safemath{\capacity}{C}						% capacity

\safemath{\normal}{\mathcal{N}}			% normal distribution
\safemath{\jpg}{\mathcal{CN}}			% jointly proper Gaussian
\safemath{\mchain}{\leftrightarrow}		% Markov chain
\safemath{\dB}{\,\mathrm{dB}}
\safemath{\dBm}{\,\mathrm{dBm}}
\safemath{\Hz}{\,\mathrm{Hz}}
\safemath{\kHz}{\,\mathrm{kHz}}
\safemath{\MHz}{\,\mathrm{MHz}}
\safemath{\GHz}{\,\mathrm{GHz}}
\safemath{\s}{\,\mathrm{s}}
\safemath{\ms}{\,\mathrm{ms}}
\safemath{\mus}{\,\mathrm{\text{\textmu}s}}
\safemath{\ns}{\,\mathrm{ns}}
\safemath{\ps}{\,\mathrm{ps}}
\safemath{\meter}{\,\mathrm{m}}
\safemath{\mm}{\,\mathrm{mm}}
\safemath{\cm}{\,\mathrm{cm}}
\safemath{\m}{\,\mathrm{m}}
\safemath{\W}{\,\mathrm{W}}
\safemath{\mW}{\, \mathrm{mW}}
\safemath{\J}{\,\mathrm{J}}
\safemath{\K}{\,\mathrm{K}}
\safemath{\bit}{\,\mathrm{bit}}
\safemath{\nat}{\,\mathrm{nat}}

\safemath{\define}{\triangleq}			% definition

\safemath{\equivalent}{\sim}
\safemath{\distas}{\sim}					% distributed according to
\safemath{\sdiff}{\Delta}				% symmetric set difference

\safemath{\reals}{\mathbb{R}}
\safemath{\positivereals}{\reals_{+}}
\safemath{\integers}{\mathbb{Z}}
\safemath{\posint}{\integers_{+}}
\safemath{\naturals}{\mathbb{N}}
\safemath{\posnaturals}{\naturals_{+}}
\safemath{\complexset}{\mathbb{C}}
\safemath{\rationals}{\mathbb{Q}}

\newcommand*{\fancyrefapplabelprefix}{app}		% Appendix
\newcommand*{\fancyrefthmlabelprefix}{thm}		% Theorem
\newcommand*{\fancyreflemlabelprefix}{lem}		% Lemma
\newcommand*{\fancyrefcorlabelprefix}{cor}		% Corollary
\newcommand*{\fancyrefdeflabelprefix}{def}		% Definition
\newcommand*{\fancyrefproplabelprefix}{prop}		% Proposition
\newcommand*{\fancyrefexmpllabelprefix}{exmpl}
\newcommand*{\fancyrefalglabelprefix}{alg}		% Algorithm
\newcommand*{\fancyreftbllabelprefix}{tbl}		% Algorithm

\frefformat{vario}{\fancyrefseclabelprefix}{Sec.~#1}
\frefformat{vario}{\fancyrefthmlabelprefix}{Thm.~#1}
\frefformat{vario}{\fancyreftbllabelprefix}{Tbl.~#1}
\frefformat{vario}{\fancyreflemlabelprefix}{Lem.~#1}
\frefformat{vario}{\fancyrefcorlabelprefix}{Cor.~#1}
\frefformat{vario}{\fancyrefdeflabelprefix}{Def.~#1}
\frefformat{vario}{\fancyreffiglabelprefix}{Fig.~#1}
\frefformat{vario}{\fancyrefapplabelprefix}{App.~#1}
\frefformat{vario}{\fancyrefeqlabelprefix}{(#1)}
\frefformat{vario}{\fancyrefproplabelprefix}{Prop.~#1}
\frefformat{vario}{\fancyrefexmpllabelprefix}{Ex.~#1}
\frefformat{vario}{\fancyrefalglabelprefix}{Alg.~#1}

%% file: macros/defs.tex
\safemath{\dictab}{[\,\dicta\,\,\dictb\,]}

\safemath{\ysig}{\bmy}
\safemath{\ysighat}{\hat{\ysig}}
\safemath{\ysigdim}{M}
\safemath{\xsig}{\bmx}
\safemath{\xsigdim}{N}
\safemath{\nx}{n_x}
\safemath{\zsig}{\bmz}
\safemath{\zsigdim}{\ysigdim}
\safemath{\rsig}{\bmr}
\safemath{\Adict}{\bA}
\safemath{\Adicttilde}{\widetilde{\Adict}}
\safemath{\Adictdim}{\outputdim\times\xsigdim}
\safemath{\avec}{\bma}
\safemath{\avectilde}{\tilde{\avec}}
\safemath{\Bdict}{\bB}
\safemath{\Bdicttilde}{\widetilde{\Bdict}}
\safemath{\Cdict}{\bC}
\safemath{\cvec}{\bmc}
\safemath{\Ddict}{\bD}
\safemath{\Ddictdim}{\ysigdim\times\xsigdim}
\safemath{\dvec}{\bmd}
\safemath{\Ddicttilde}{\widetilde{\bD}}
\safemath{\Bonb}{\bB}
\safemath{\bvec}{\bmb}
\safemath{\Bonbdim}{\ysigdim\times\ysigdim}
\safemath{\noise}{\bmn}
\safemath{\noisedim}{\ysigim}
\safemath{\err}{\bme}
\safemath{\errdim}{\ysigdim}
\safemath{\errset}{\setE}
\safemath{\nerr}{n_e}
\safemath{\delop}{\bP_\errset}
\safemath{\delopc}{\bP_{{\errset}^c}}

\safemath{\cplxi}{\imath}
\safemath{\cplxj}{\jmath}
\safemath{\dict}{\matD}
\safemath{\inputdim}{N}		% number of columns of dictionary D
\safemath{\outputdim}{M}		%number of rows of dictionary D
\safemath{\sparsity}{S}	%sparsity
\safemath{\inputdimA}{{N_a}}	%total number of elements in dictionary A
\safemath{\inputdimB}{{N_b}}	%total number of elements in dictionary B
\safemath{\elemA}{{n_a}}	%number of elements chosen from dictionary A
\safemath{\elemB}{{n_b}}	%number of elements chosen from dictionary B
\safemath{\resA}{\matR_a}	%restriction map to elements of dictionary A
\safemath{\resB}{\matR_b}	%restriction map to elements of dictionary B
\safemath{\subD}{\matS} %subdictionary
\safemath{\subA}{\matS_a} %subdictionary part of A
\safemath{\subB}{\matS_b} %subdictionary part of B
\safemath{\dicta}{\matA} 	% first subdictionary
\safemath{\dictb}{\matB} 	% second subdictionary
\safemath{\hollowS}{H}
\safemath{\hollowA}{H_a}
\safemath{\hollowB}{H_b}
\safemath{\cross}{Z}
\safemath{\coh}{\mu_d}			% coherence dictionary
\safemath{\coha}{\mu_a}			% coherence first subdictionary
\safemath{\cohb}{\mu_b}			% coherence second subdictionary
\safemath{\mubs}{\nu}	%block sub-coherence
\safemath{\cohm}{\mu_m} %mutual coherence
\safemath{\dictset}{\setD}	% set of dictionaries
\safemath{\dictsetp}{\dictset(\coh,\coha,\cohb)}	% set of dictionaries parametrized
\safemath{\dictsetgen}{\dictset_\text{gen}}
\safemath{\dictsetgenp}{\dictsetgen(\coh)}
\safemath{\dictsetonb}{\dictset_\text{onb}}
\safemath{\dictsetonbp}{\dictsetonb(\coh)}

\safemath{\leftside}{U}
\safemath{\rightsideA}{R_a}
\safemath{\rightsideB}{R_b}

\safemath{\indexS}{\setI_S} %set of indices participating in sub-dictionary S

\safemath{\na}{n_a}			% cardinality of set of linearly independent columns of first dictionary
\safemath{\nb}{n_b}			% cardinality of set of linearly independent columns of second dictionary
\safemath{\coeffa}{p_i}	%coefficients for columns of A
\safemath{\coeffb}{q_j}	%coefficients for columns of B
\safemath{\seta}{\setP}		% set of linearly independent columns of A
\safemath{\setb}{\setQ}     % set of linearly independent columns of B
\safemath{\setw}{\setW}	%set of n largest elements of w
\safemath{\setz}{\setZ}	%set of L-n largest elements of z
\safemath{\cola}{\veca}		% generic element of the dictionary A
\safemath{\colb}{\vecb}		% generic element of the dictionary B
\safemath{\cold}{\vecd}		% generic element of the dictionary D
\safemath{\inputvec}{\vecx} 	%coefficient vector (input)
\safemath{\error}{\vece}	%error vector
\safemath{\noiseout}{\vecz} 	%noisy output vector
\safemath{\inputvecel}{x}
\safemath{\inputveca}{\vecx_a}
\safemath{\inputvecb}{\vecx_b}
\safemath{\outputvec}{\vecy}	%output of Dictionary
\safemath{\lambdamin}{\lambda_{\mathrm{min}}}
\safemath{\elltwo}{\ell_2}
\safemath{\ellone}{\ell_1}
\safemath{\ellzero}{\ell_0}
\safemath{\ellinf}{\ell_\infty}
\safemath{\ellinftilde}{\ell_{\widetilde\infty}}
\safemath{\licard}{Z(\coh,\coha,\cohb)}
\safemath{\xsol}{\hat{x}}
\safemath{\xbord}{x_b}		%Solution at the border
\safemath{\xstat}{x_s}		%Solution stationary in l0 prob
\safemath{\xstatLone}{\tilde{x}_s}
\safemath{\order}{\mathcal{O}} %order notation (big O)
\safemath{\scales}{\Theta} %scales as
\safemath{\ones}{\mathbf{1}} %all ones matrix
\safemath{\zeroes}{\mathbf{0}} %all zeroes matrix
\safemath{\thlone}{\kappa(\coh,\cohb)} %treshold l1 problem
\safemath{\constoneA}{\delta} %constant in l1 theorem to save space
\safemath{\constoneB}{\epsilon} %constant in l1 theorem to save space
\safemath{\nlarge}{L}				   %num large elements
\safemath{\sumlarge}{S_\nlarge}
\safemath{\maxlarger}{P_\nlarge}	   % maximum in Gribonval and Nielsen
\safemath{\Pzero}{\textrm{P0}}	
\safemath{\Pone}{\textrm{P1}}
\safemath{\vecfir}{\vecw}			 % \vecv element of the kernel of the dictionary, \vecv=[\vecfir \vecsec]
\safemath{\vecsec}{\vecz}
\safemath{\elvecfir}{w}              % element of vecfir
\safemath{\elvecsec}{z}				 % element of vecsec
\safemath{\nlargefir}{n}
\safemath{\normout}{\gamma}
\safemath{\auxfun}{h}
\safemath{\supp}{\textrm{supp}}%support

\safemath{\indexa}{\ell}
\safemath{\indexb}{r}
\safemath{\indexc}{i}
\safemath{\indexd}{j}

\safemath{\project}{P}%projector

%% file: sections/0-abstract.tex
% !TEX root = weather_sensing_paper.tex
% DO NOT REMOVE THE ABOVE COMMENT!

\begin{abstract}
One of the key features of \gls{6g} mobile communications will be \gls{isac}. While the main goal of \gls{isac} in standardization efforts is to detect objects, the byproducts of radar operations can be used to enable new services in \gls{6g}, such as weather sensing. Even though weather radars are the most prominent technology for weather detection and monitoring, they are expensive and usually neglect areas in close vicinity. To this end, we propose reusing the dense deployment of 6G \acrlongpl{bs} for weather sensing purposes by detecting and estimating weather conditions. We implement both a classifier and a regressor as a \acrlong{cnn} trained across measurements with varying precipitation rates and wind speeds. We implement our approach in an \gls{isac} \acrlong{poc} and conduct a multi-week experiment campaign. Experimental results show that we are able to jointly and accurately classify weather conditions with accuracies of $99.38\%$ and $98.99\%$ for precipitation rate and wind speed, respectively. For estimation, we obtain mean absolute errors of $1.2$ mm/h and $1.5$ km/h, for precipitation rate and wind speed, respectively. These findings indicate that weather sensing services can be reliably deployed in \gls{6g} \gls{isac} networks, broadening their service portfolio and boosting their market value.
\end{abstract}

%% file: sections/1-introduction.tex
% !TEX root = weather_sensing_paper.tex
% DO NOT REMOVE THE ABOVE COMMENT!
\glsresetall

\section{Introduction} 
\label{sec:introduction}
One of the most prominent and distinctive features of \gls{6g} mobile communication systems will be \gls{isac}. This technology leverages existing communication infrastructure to perform tasks beyond data transmission, such as drone detection, intruder detection, as well as traffic monitoring \cite{Mandelli2023}. Unlike in positioning technologies, where targets actively send and receive signals, sensing targets remain entirely passive. Beyond the aforementioned target detection-focused use cases, \gls{isac} can also be used to monitor weather conditions in the cellular network coverage.

\Gls{em} waves are affected by atmospheric conditions, e.g., clouds and precipitation, through reflection, absorption, scattering, refraction, and diffraction \cite{Oguchi1983}. These effects are exploited by weather radars, which are the primary deployed technology for weather detection, monitoring, and prediction. These radars emit \gls{em} pulses which are partially reflected back to the radar due to the presence of scatterers, and by measuring the propagation time, the distances to the scatterers can be determined. Additionally, the received power provides insights into the size and density of the scatterers. Furthermore, to analyze the motion and speed of weather phenomena, weather radars exploit the Doppler effect. Despite their accuracy in determining weather conditions over large areas, weather radars are expensive, resulting in their sparse deployment. Moreover, since weather radars are configured to achieve full coverage by mechanically sweeping a limited number of elevation angles over a predefined time, areas in the close vicinity of the radar can be overlooked due to scan timing and revisit constraints \cite{Palmer2023}. In contrast, \gls{isac} networks are densely deployed and also use \gls{em} waves, making them a promising complement or alternative for weather sensing.
\subsection{Contributions}
To address weather detection and monitoring at close range, we propose a solution for enabling weather sensing in \gls{6g} \gls{isac} networks. In this context, we operate a communication system in a radar-style approach with half-duplex \glspl{ru}. By collecting the transmitted and received signals, we estimate the \gls{csi}. From the estimated \gls{csi}, we compute features to be used as input data for our weather estimation supervised learning model. In the presented problem, we perform two different tasks, classification and regression, in order to jointly classify or estimate both precipitation rate and wind speed. We demonstrate the efficacy of our method through experiments with a pre-existing FR2 \gls{isac} \gls{poc} \cite{Wild2023}. We collect training and test data over multiple weeks in scenarios with and without rain and with variable precipitation rates and wind speeds. For training of the supervised model, we collect reference data from a weather station \cite{weatherstation} in close vicinity to the \gls{poc} (see \fref{sec:weather_station}). 
\subsection{Related Work}
Recent works on weather radar include \cite{Yu2020,Wang2020,Villa2024}. Both \cite{Yu2020} and \cite{Wang2020} use the interferometric phase diagram as input to classify rain vs. no rain. Additionally, \cite{Yu2020} incorporates clutter phase alignment and Doppler velocity in its methodology. In contrast, \cite{Villa2024} focuses on tracking weather evolution over time, including the course and speed of weather phenomena. Unlike these studies, our approach leverages capabilities of future {\gls{6g}} \gls{isac} networks to estimate weather conditions, not only using existing hardware, but also without relying on traditional weather radars.

Most works on weather monitoring with telecommunication hardware exploit microwave \cite{Lian2022,Daher2022,Jacoby2024,Blettner2023,Overeem2021,Djibo2023,Diba2021} or \gls{mmwave} \cite{Han2019} links, typically point-to-point, backhaul between \acrlongpl{bs}, or \gls{los} transmission links to extract rain measurements. These studies mainly address rain vs. no rain classification \cite{Lian2022} or precipitation rate estimation \cite{Lian2022,Daher2022,Han2019,Jacoby2024,Diba2021} by analyzing signal attenuation on static links rather than the full complex-valued \gls{csi}.

Methods used for rain classification and estimation include \acrlongpl{slp} \cite{Daher2022}, \acrlongpl{svm} \cite{Lian2022}, \glspl{cnn} \cite{Lian2022}, \acrlongpl{rnn} \cite{Jacoby2024}, \acrlongpl{lstm} \cite{Lian2022,Diba2021}, and \acrlongpl{gru} \cite{Lian2022}. Some studies also generate rainfall maps \cite{Djibo2023,Blettner2023,Overeem2021}. In contrast, our method does not rely on static microwave links. Instead, we utilize existing cellular network infrastructure used for communication between \glspl{gnbru} and \glspl{ue}. While we also estimate precipitation rates, our approach uses \glspl{cnn} to process \gls{csi}-based periodograms, which are range–Doppler shift images, enabling joint estimation of precipitation rate and wind speed. Moreover, prior methods rely on signal attenuation solely from unobstructed fixed links where \gls{los} is assumed, i.e., they are not robust to cluttered environments with multiple obstructions and reflections. 

%% file: sections/2-periodogram_generation.tex
% !TEX root = weather_sensing_paper.tex
% DO NOT REMOVE THE ABOVE COMMENT!

\section{Data Pre-Processing} 
\label{sec:data_pre_procesing}
In this section, we detail the data pre-processing pipeline, from the transmission and reception of \gls{ofdm} frames to the computation of weather features.
\subsection{System Description} 
\label{sec:system_model}
We consider a system with a \gls{tx} and a \gls{rx} that operates in half-duplex mode, i.e., can both transmit and receive signals, but not simultaneously. The \gls{tx} transmits $S$ frames with $M$ \gls{ofdm} symbols at carrier frequency $f_c$ over $N$ subcarriers spaced by $\Delta f$. We assume analog beamforming, i.e., one active beam at a time. While these transmitted signals propagate to the \glspl{ue}, they are also reflected by the environment back to the \gls{rx}. We define the \gls{tx} and \gls{rx} \gls{ofdm} frames as $\bT_s, \, \bR_s \in \opC^{N \times M}$, for $s \in \setS=\chav{1,\dots,S}$. In the matrices $\bT_s$ and $\bR_s$, the rows and columns represent the subcarriers and OFDM symbols, respectively.
\subsection{CSI Estimation}
To estimate weather conditions, we must first provide an estimate of the \gls{csi} matrix. The \gls{csi} matrix describes the signal propagation between \gls{tx} and \gls{rx} and captures reflections from multiple scatterers such as buildings, people, drones, and cars. 
Depending on the \gls{isac} use case, these scatterers might be considered targets or clutter components. Moreover, the \gls{csi} matrix also encodes weather phenomena due to {multiple effects, such as attenuation and reflections} caused by weather conditions. These scatterers and weather phenomena affect the \gls{tx} \gls{ofdm} frame by introducing Doppler shifts and propagation delays within the channel between \gls{tx} and \gls{rx}. For sensing purposes, the \gls{tx} \gls{ofdm} frame $\bT_s$ serves only as a reference signal. We estimate the channel by the element-wise division of the \gls{rx} \gls{ofdm} frame $\bR_s$ by the \gls{tx} \gls{ofdm} frame $\bT_s$ as
\begin{align} \label{eq:csi_matrix}
\PR{\bH_s}_{k,l} = \frac{\PR{\bR_s}_{k,l}}{\PR{\bT_s}_{k,l}},
\end{align}
where $\bH_s \in \opC^{N \times M}$ is the \gls{csi} matrix produced by the $s$th \gls{tx} and \gls{rx} frames, $k$ is the row index, and $l$ is the column index. To enable efficient \glspl{fft} in subsequent weather-feature computation, we zero-pad the \gls{csi} matrix $\bH_s \in \opC^{N \times M}$ to the next power of $2$, giving us $\bar{\bH}_s \in \opC^{N_{\text{pad}} \times M_{\text{pad}}}$, where $N_{\text{pad}} = 2^{\ceil{\log_2 N}}$ and $M_{\text{pad}} = 2^{\ceil{\log_2 M}}$.
\subsection{Clutter Removal} 
\label{sec:clutter_removal}
We highlight that in the case of weather sensing, radio resources might not be available to acquire measurements from an empty environment without capturing reflections from other scatterers. This environment would be the most suitable for weather detection and estimation, but it is less suitable for communications and other sensing use cases. Thus, to mitigate the effects of clutter, i.e., unwanted scatterers, we apply clutter removal to the zero-padded \gls{csi} matrix~$\bar{\bH}_s$, even though not strictly required for weather sensing. 

The goal of clutter removal is to suppress environmental components irrelevant to the sensing task. For weather detection and estimation, we are solely interested in analyzing the effects of raindrops, dust particles, and snowflakes, while attenuating the contributions from the surrounding environment (e.g., buildings). For clutter removal, we use the \gls{crap} algorithm \cite{Henninger2023}. We refer to the \gls{csi} matrix after \gls{crap} as $\hat{\bH}_s \in \opC^{N_{\text{pad}} \times M_{\text{pad}}}$. This algorithm requires a calibration matrix, which we compute based on measurements of the previous day with the lowest precipitation rate and wind speed.
\begin{figure}[tp]
    \centering
    \includegraphics[width=\columnwidth]{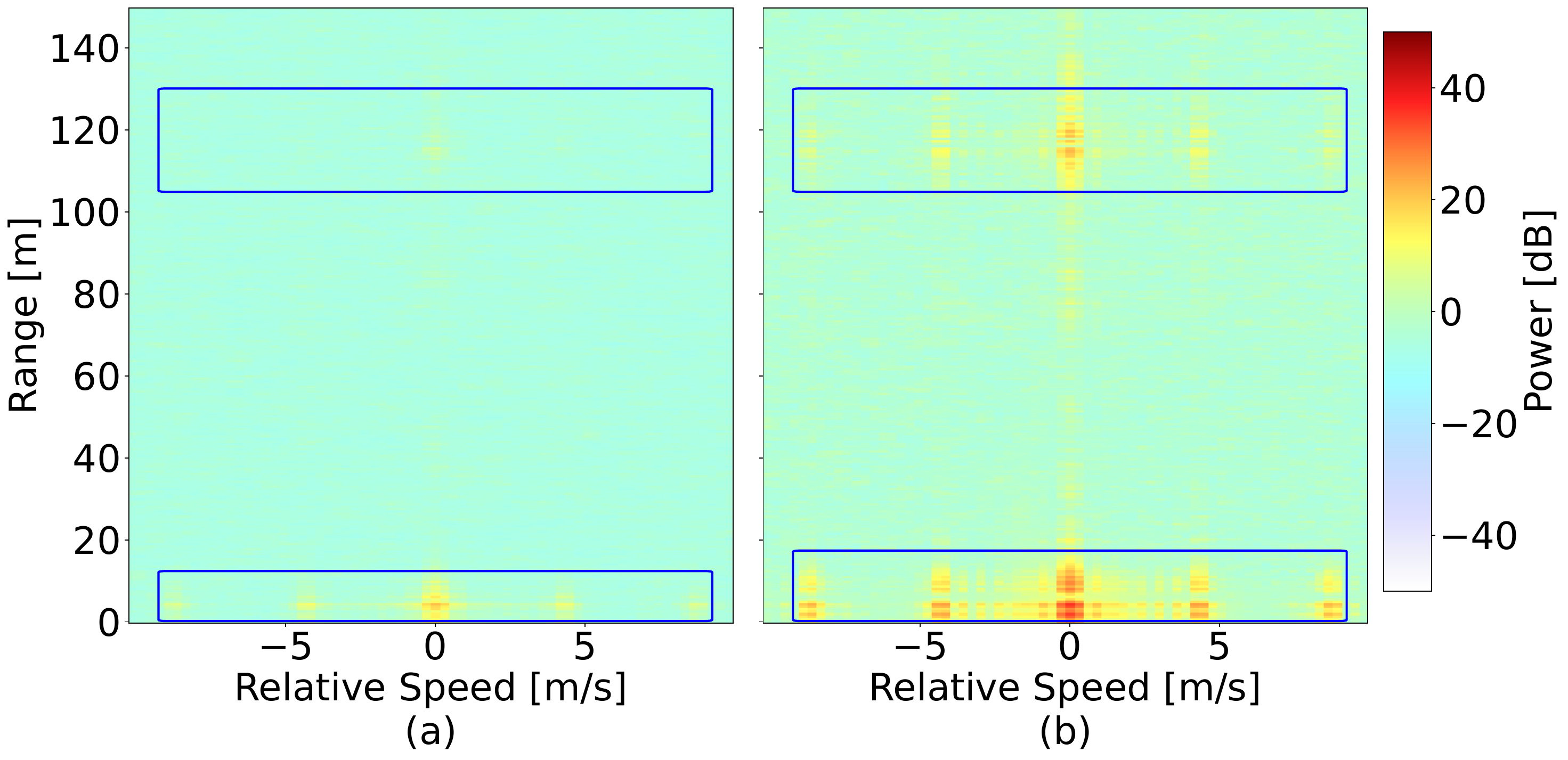}
    \caption{Comparison between periodograms (a) without rain and (b) with rain. We observe that the rain increases the noise floor of the periodogram and emphasizes the presence of targets around $0$ and $120$ meters from the \gls{rx}.}
    \label{fig:periodograms}
\end{figure}
\subsection{OFDM Radar Processing}
The next step towards weather detection and estimation is periodogram generation. A periodogram estimates a signal’s power spectral density, highlighting its dominant spectral components, and is computed via {a series of} \glspl{fft}. In the case of this work, we aim to compute a \gls{2d} periodogram that enables us to detect targets at various ranges (delays) and Doppler shifts (relative speeds) {\cite{Braun2014}}. Considering our analog beamforming assumption and no beam sweeping, in this work, we disregard antenna and angular information. More specifically, we focus on analyzing the effect of weather elements such as raindrops, dust particles, and snowflakes in the periodogram. Since the input of the periodogram computation is a \gls{2d} \gls{csi} matrix, we extend the periodogram calculation as follows:
\begin{align} \label{eq:complex_periodogram}
\PR{\hat{\bP}_s}_{n,m} = &\frac{1}{N M} \sum_{k=0}^{N_{\text{pad}}-1} \PR{\sum_{l=0}^{M_{\text{pad}}-1} \PR{\hat{\bH}_s}_{k,l} e^{-j2\pi \frac{lm}{M_{\text{pad}}}}} e^{j2\pi \frac{kn}{N_{\text{pad}}}}\nonumber\\
= &\frac{1}{N M} \setF^{-1}\chav{\setF\chav{\PR{\hat{\bH}_s}_{k,l}}}.
\end{align}
In \fref{eq:complex_periodogram}, we perform $N$ \glspl{fft} of length $M_{\text{pad}}$ that take the \gls{csi} matrix from time (\gls{ofdm} symbols) to Doppler shift (relative speed) domain and $M$ \glspl{ifft} of length $N_{\text{pad}}$ that take the \gls{csi} matrix from frequency (subcarriers) to delay (range) domain. We crop the periodogram $\hat{\bP}_s \in \opC^{N_{\text{pad}} \times M_{\text{pad}}}$ to $\PR{1,\dots,N'}$ rows and $\PR{\floor{\PR{M_{\text{pad}}-M'}/2}+1,\dots,\floor{\PR{M_{\text{pad}}-M'}/2}+M'}$ columns according to the desired evaluation range, i.e., the interval of ranges and relative speeds that we are interested in. To decrease computational complexity, cropping parameters can be constrained, though at the cost of limited detection range/speed. We highlight that in the case of relative speed, the {cropping} is centered around zero. The calculation of $N'$ range bins and $M'$ speed bins can be carried out according to \cite{Braun2014}. In this way, from a \gls{2d} \gls{csi} matrix $\bH_s \in \opC^{N \times M}$ over subcarriers and \gls{ofdm} symbols, we obtain a \gls{2d} complex periodogram $\bP_s \in \opC^{N' \times M'}$ over range (delay) and Doppler shift (relative~speed). 
\subsection{Feature Engineering} 
\label{sec:feature_engineering}
In \fref{fig:periodograms}, we present the absolute-value squared periodograms $|\bP_s|^2$ for scenarios without and with rain, where the rain-induced effects are clearly visible, as highlighted by the blue rectangles in the figure. Nevertheless, we highlight that wind detection relies on reflective particles in the air, such as raindrops, dust particles, or snowflakes. In this work, we propose using the \gls{2d} periodograms as input data to compute the features of the weather detection and estimation problem. 

In this work, we consider that the \gls{tx} emits waves in polarization $\rho_1$ and that the \gls{rx} receives waves both in polarization $\rho_1$ and $\rho_2$. Therefore, we can generate complex periodograms for two polarization combinations: $\rho_1$-$\rho_1$ ($\bP_s^{\rho_1\rho_1}$) and $\rho_1$-$\rho_2$ ($\bP_s^{\rho_1\rho_2}$). In the first case, the CSI matrix is computed using \gls{tx} and \gls{rx} \gls{ofdm} frames from polarization $\rho_1$. In the second case, the CSI matrix is computed using \gls{tx} \gls{ofdm} frames from polarization $\rho_1$ and \gls{rx} \gls{ofdm} frames from polarization $\rho_2$. Analyzing both polarizations provides additional information, e.g., by revealing distinct clutter characteristics. In addition, we decompose the complex periodogram into its real and imaginary parts. Thus, we have the following \gls{3d} feature tensor $\bF_s \in \opR^{N' \times M' \times L}$ with $L=4$ channels:
\begin{align}
    \PR{\bF_s}_{n,m,1} &= \Re\chav{\PR{\bP_s^{\rho_1\rho_1}}_{n,m}}\\ \PR{\bF_s}_{n,m,2} &= \Im\chav{\PR{\bP_s^{\rho_1\rho_1}}_{n,m}}\\
    \PR{\bF_s}_{n,m,3} &= \Re\chav{\PR{\bP_s^{\rho_1\rho_2}}_{n,m}}\\
    \PR{\bF_s}_{n,m,4} &= \Im\chav{\PR{\bP_s^{\rho_1\rho_2}}_{n,m}},
\end{align}
where $n$ and $m$ are the indices of the first and second dimensions of the tensor, respectively. After computing the \gls{3d} feature tensor $\bF_s$ from each \gls{csi} matrix $\bH_s$, we normalize the matrices so as to improve convergence on the weather detection and estimation problem. Specifically, we perform a channel-wise normalization: for each channel $\ell = 1, \dots, L$, the data is normalized across all samples $S$, $N'$ rows, and $M'$ columns, such that each channel has zero mean and unit standard deviation. We define the normalized \gls{3d} feature tensor as $\bar{\bF}_s$. The set of tensors $\chav{\bar{\bF}_s}_{s \in \setS}$ forms the feature dataset. 

%% file: sections/3-weather_sensing.tex
% !TEX root = weather_sensing_paper.tex
% DO NOT REMOVE THE ABOVE COMMENT!

\section{Weather Estimation} 
\label{sec:weather_sensing}
\begin{figure*}[t]
    \centering
    \includegraphics[width=0.8\textwidth]{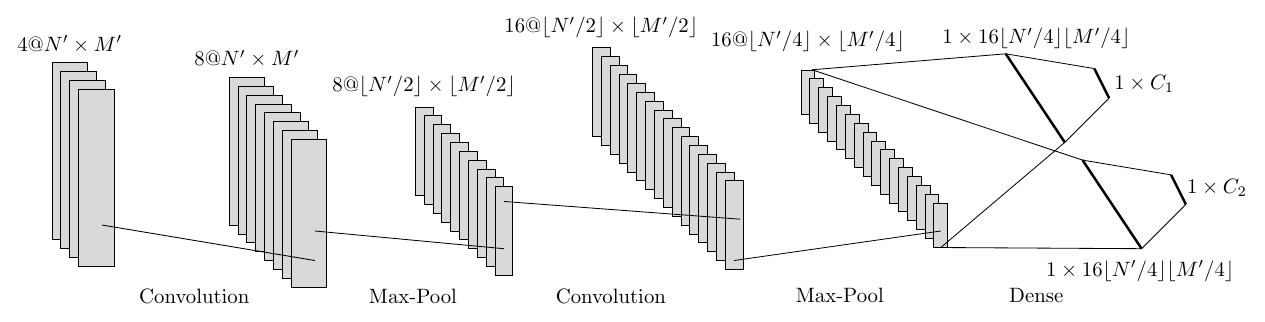}
    \caption{CNN architecture for the weather detection and estimation problem.}
    \label{fig:CNN}
\end{figure*}
We now present the weather detection and estimation problem through two different tasks: classification and regression. The goal of these tasks is to jointly detect and estimate $W=2$ weather metrics $w \in \{1,2\}$: precipitation rate ($w=1$) and wind speed ($w=2$). For these tasks, we can learn a function $g_{\boldsymbol\theta}: \opR^{N' \times M' \times L} \rightarrow \opR^{C_w}$ that maps feature tensors to estimates of weather metrics $\hat{\bmx}_{s,w} \in \opR^{C_w}$, for $w\in \chav{1,2}$, as
\begin{align} \label{eq:learning_function}
    \hat{\bmx}_{s,w} = g_{\boldsymbol\theta}\PC{\bar{\bF}_s} \in \opR^{C_w} \text{ for } w\in \chav{1,2},
\end{align}
where $C_w$ is the number of classes of the specific weather metric $w$. We implement $g_{\boldsymbol\theta}$ as a \gls{cnn} with weights and biases described by $\boldsymbol\theta$.
\subsection{Task I: Classification} 
\label{sec:classification}
A classification problem entails learning a mapping from features to a finite set of discrete classes. We learn the classification function  $g_{\boldsymbol\theta}$ with a cross-entropy loss as
\begin{align} \label{eq:loss_classification}
    \loss_{\text{class}} = \sum_{w=1}^W \sum_{s=1}^S -\phi_{w,y} \log \PR{\frac{\exp{\PC{\PR{\hat{\bmx}_{s,w}}_y}}}{\sum_{j=1}^{C_w} \exp{\PC{\PR{\hat{\bmx}_{s,w}}_j}}}},
\end{align}
where $\phi_{w,y}$ is the weight of the class $y$ in weather metric $w$, $\PR{\hat{\bmx}_{s,w}}_y$ is the output of the CNN for sample $s$ in weather metric $w$ and class $y$, and class $y=y_{s,w}$ is the correct label for frame $s$ and weather metric $w$. The predicted label for frame $s$ and weather metric $w$ is then
\begin{align} \label{eq:estimated_class}
    \hat{y}_{s,w} = \underset{j=1,\dots,C_w}{\argmax} \PR{\hat{\bmx}_{s,w}}_j.
\end{align}
\subsection{Task II: Regression} \label{sec:regression}
A regression problem entails estimating a continuous value based on the input data. Since we have no classes in this case, we assume that $C_w=1$, for $w \in \chav{1,2}$. We learn the regression function  $g_{\boldsymbol\theta}$ with a \gls{mse} loss~as
\begin{align} \label{eq:loss_regression}
    \loss_{\text{reg}} = \sum_{w=1}^W \frac{1}{S} \sum_{s=1}^S \PM{\hat{x}_{s,w}-y_{s,w}}^2,
\end{align}
where $\hat{x}_{s,w}$ is the output of the CNN and $y_{s,w}$ is the reference data, both for sample $s$ and weather metric $w$. The predicted label for sample $s$ and weather metric $w$ is then 
\begin{align} \label{eq:estimated_value}
\hat{y}_{s,w} =\hat{x}_{s,w}.
\end{align}
Both loss functions in \fref{eq:loss_classification} and \fref{eq:loss_regression} rely on reference data for all training samples, i.e., it is a supervised learning problem. 
\subsection{CNN Architecture}
As illustrated in \fref{fig:CNN}, $g_{\boldsymbol\theta}$ is implemented as a \gls{cnn} with the following structure:
\begin{itemize}
    \item Convolutional layer with $4$ input channels, $8$ output channels, kernel size of $3\times3$, stride of $1$, and padding of $1$. 
    \item Max pooling layer with kernel size of $2\times2$, stride of $2$, and padding of $0$. 
    \item Convolutional layer with $8$ input channels, $16$ output channels, kernel size of $3\times3$, stride of $1$, and padding of $1$. 
    \item Max pooling layer with kernel size of $2\times2$, stride of $2$, and padding of $0$. 
    \item Flatten layer that vectorizes the \gls{3d} output of the max pooling layer into a \acrlong{1d} vector. 
    \item Two independent fully connected layers with input size of $16\floor{N'/4}\floor{M'/4}$ and output size $C_w$, for $w \in \chav{1,2}$.
\end{itemize}
These hyperparameters were chosen heuristically after multiple experiments with a pre-trained ResNet-18 \cite{He2016} and custom CNNs built from scratch with varying numbers of convolutional and pooling layers, channel sizes (e.g., 32, 64, 96, 128), and kernel sizes (e.g., 5, 6). The final selection was based on computational complexity and the resulting training and validation losses.

%% file: sections/4-results.tex
% !TEX root = weather_sensing_paper.tex
% DO NOT REMOVE THE ABOVE COMMENT!

\section{Experimental Results} 
\label{sec:results}
In this section, we provide details about the {pre-existing FR2} \gls{isac} \gls{poc} and describe the collection of weather data used in the supervised learning problems. Furthermore, we describe the performance metrics used and showcase the effectiveness of the proposed method through experimental results.
\subsection{FR2 ISAC PoC Setup}
\begin{figure}
 \begin{subfigure}[t]{0.3\columnwidth}
     \includegraphics[height=4.1cm]{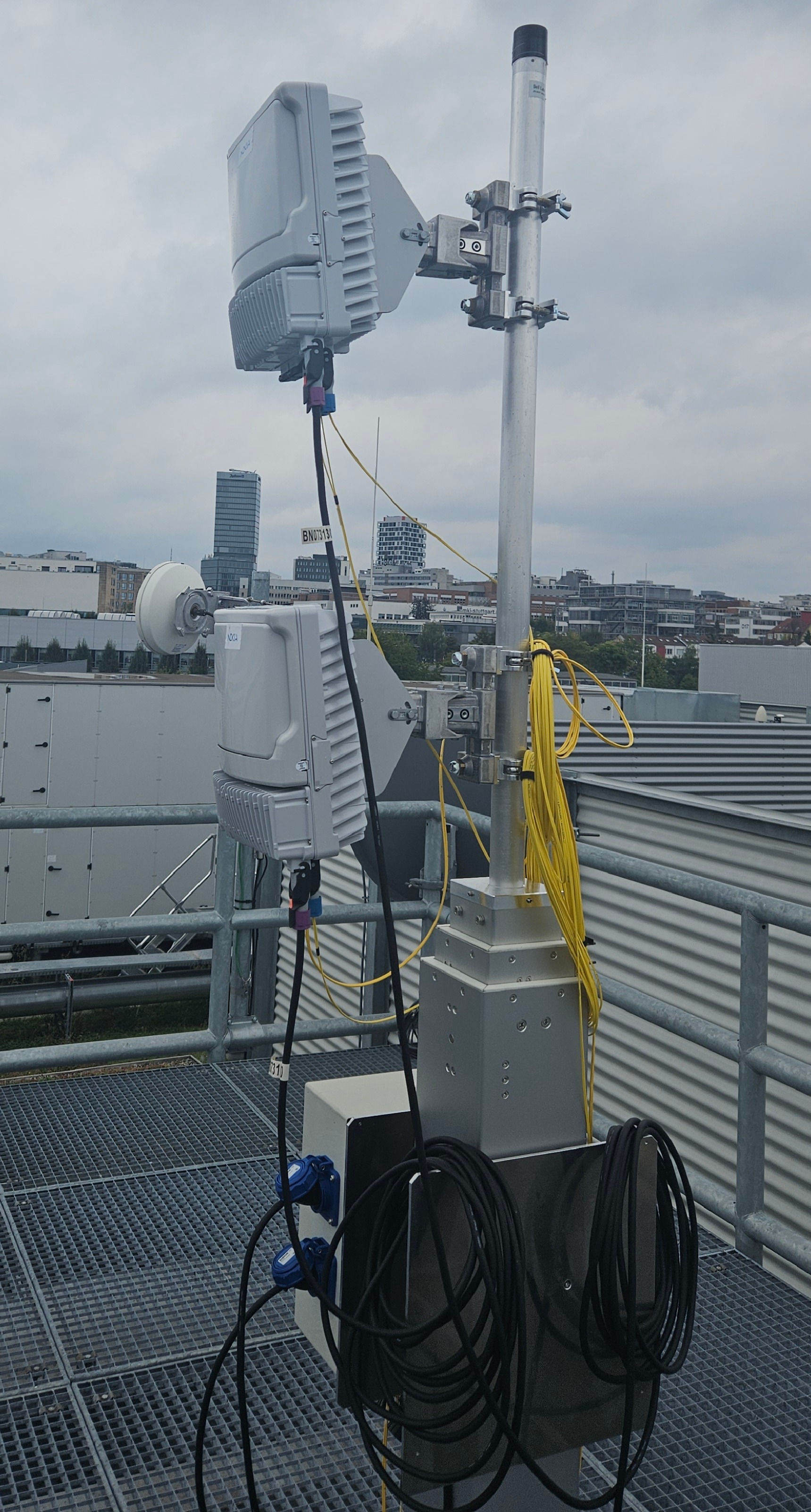}
     \caption{\glspl{ru}}
     \label{fig:rus}
 \end{subfigure}%
 \hfill
 \begin{subfigure}[t]{0.4\columnwidth}
     \includegraphics[height=4.1cm]{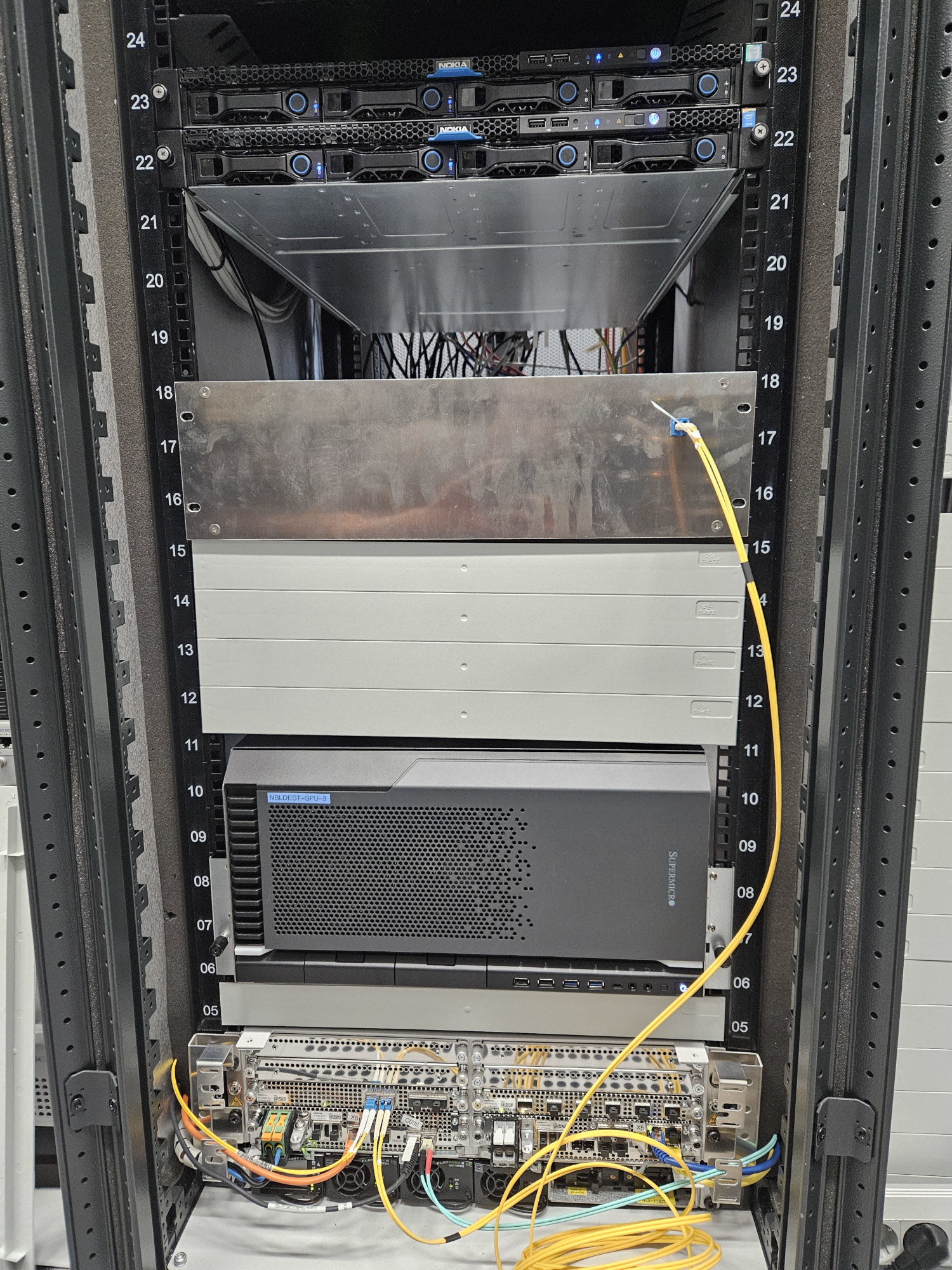}
     \caption{\gls{spu} and synchronization hardware}
     \label{fig:spu}
 \end{subfigure}%
 \hfill
 \begin{subfigure}[t]{0.2\columnwidth}
     \includegraphics[height=4.1cm]{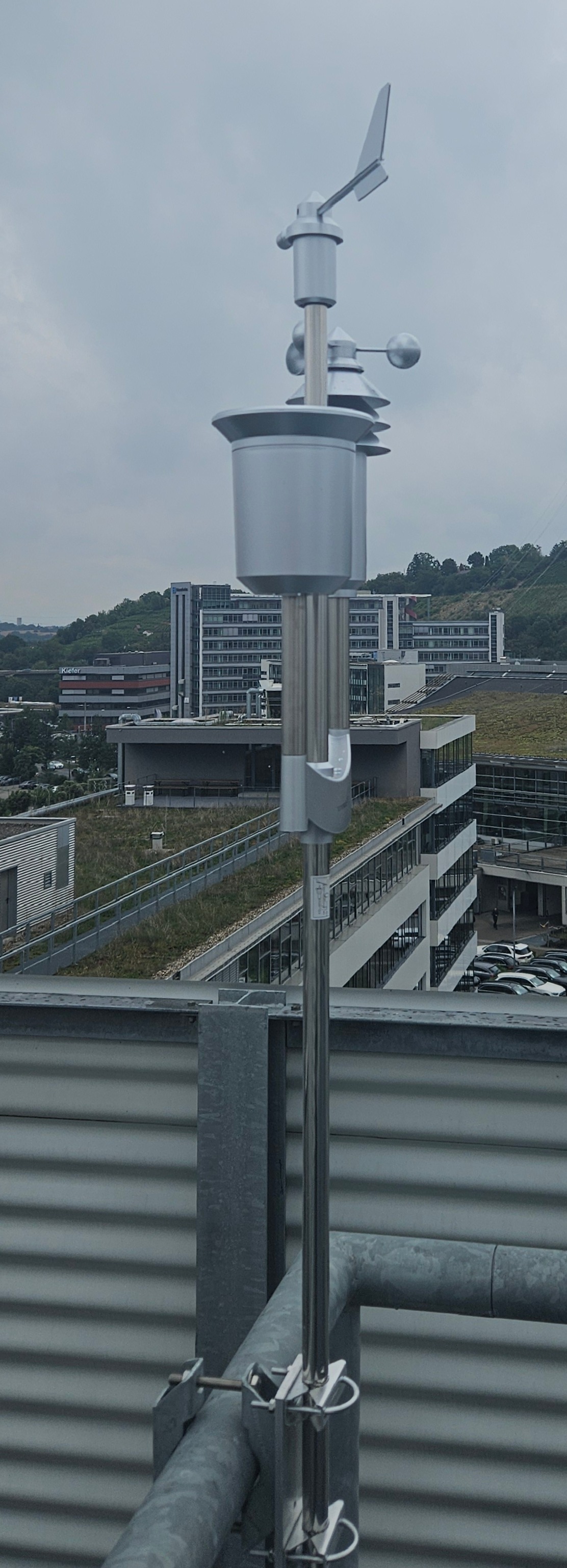}
     \caption{Weather station}
     \label{fig:weather_station}
 \end{subfigure}%
 \caption{\gls{isac} PoC deployment for weather sensing experiments.}
 \label{fig:poc_photos}
\end{figure}

Figs. \ref{fig:rus}, \ref{fig:spu} show the deployment of the FR2 \gls{isac} \gls{poc}, which consists~of:
\begin{itemize}
\item $2$ \gls{ru}s, a \gls{gnbru} acting as a \gls{tx} and a sniffer acting as \gls{rx} (see \fref{fig:rus}). Each \gls{ru} consists of 2 single-polarized (vertical  and horizontal, $\rho_1$ and $\rho_2$, as defined in \fref{sec:feature_engineering}) \acrlongpl{ura}, consisting of $8 \times 12$ antenna elements each. Transmission on a single polarization allows up to $52$\,dBm \gls{eirp}. The \gls{ru}s support analog beamforming, with the FR2 \gls{isac} \gls{poc} allowing for beam switching at each radio frame.
\item a server responsible for sensing processing, the \gls{spu} (see \fref{fig:spu}), including periodogram computation and weather detection and estimation.
\item a synchronization source (see \fref{fig:spu}) for the antennas and server.
\end{itemize}
The \gls{tx} and \gls{rx} antennas are collocated in a quasi mono-static configuration. The deployment location of the antennas is the roof of a 5 story office building within a city. The antennas are \gls{5g} mmWave \glspl{ru} for \gls{tx} and \gls{rx}, conforming to the \gls{ecpri} split 7.2x\cite{oran_wg4_spec_v19}. The \gls{poc} uses \gls{5g} $\mu=3$ numerology, typical for FR2. Each radio frame consists of $M=1120$ \gls{ofdm} symbols over $N=1584$ subcarriers spaced by $\Delta f = 120$ KHz within a bandwidth of 200 MHz. We use a center frequency of $f_c = 27.6$ GHz, as available in our over-the-air license for our outdoor deployment. For the \gls{poc}, we use a pre-recorded \gls{ofdm} test frame, which is continuously transmitted over the fronthaul link from the \gls{spu} to the \gls{gnbru}. Each subcarrier in the test frame is modulated with a random QPSK constellation point. This test frame is emitted by the \gls{gnbru} towards the environment and the signal reflected by the environment is then received by the sniffer, transmitted over fronthaul to the \gls{spu}, which then performs \gls{csi} and periodogram computation, as described in \fref{sec:data_pre_procesing}. Further details on the \gls{poc} are available in \cite{Wild2023}.
\begin{figure}[t]
\centering
\includegraphics[width=0.9\columnwidth]{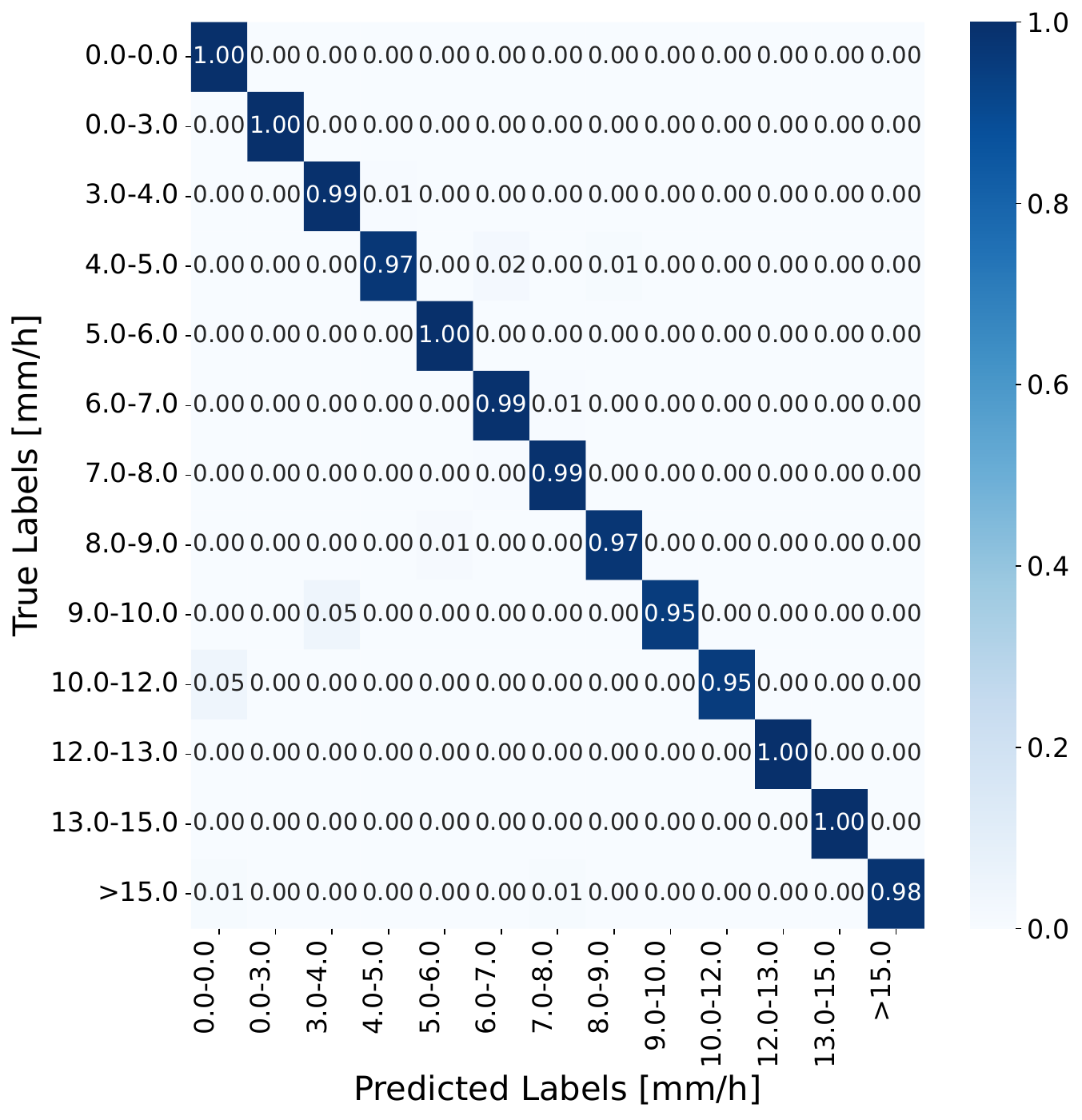}
    \caption{Precipitation rate classification accuracy on the test set.}
    \label{fig:rain_accuracy}
\end{figure}
\begin{figure}
\centering
    \includegraphics[width=0.85\columnwidth]{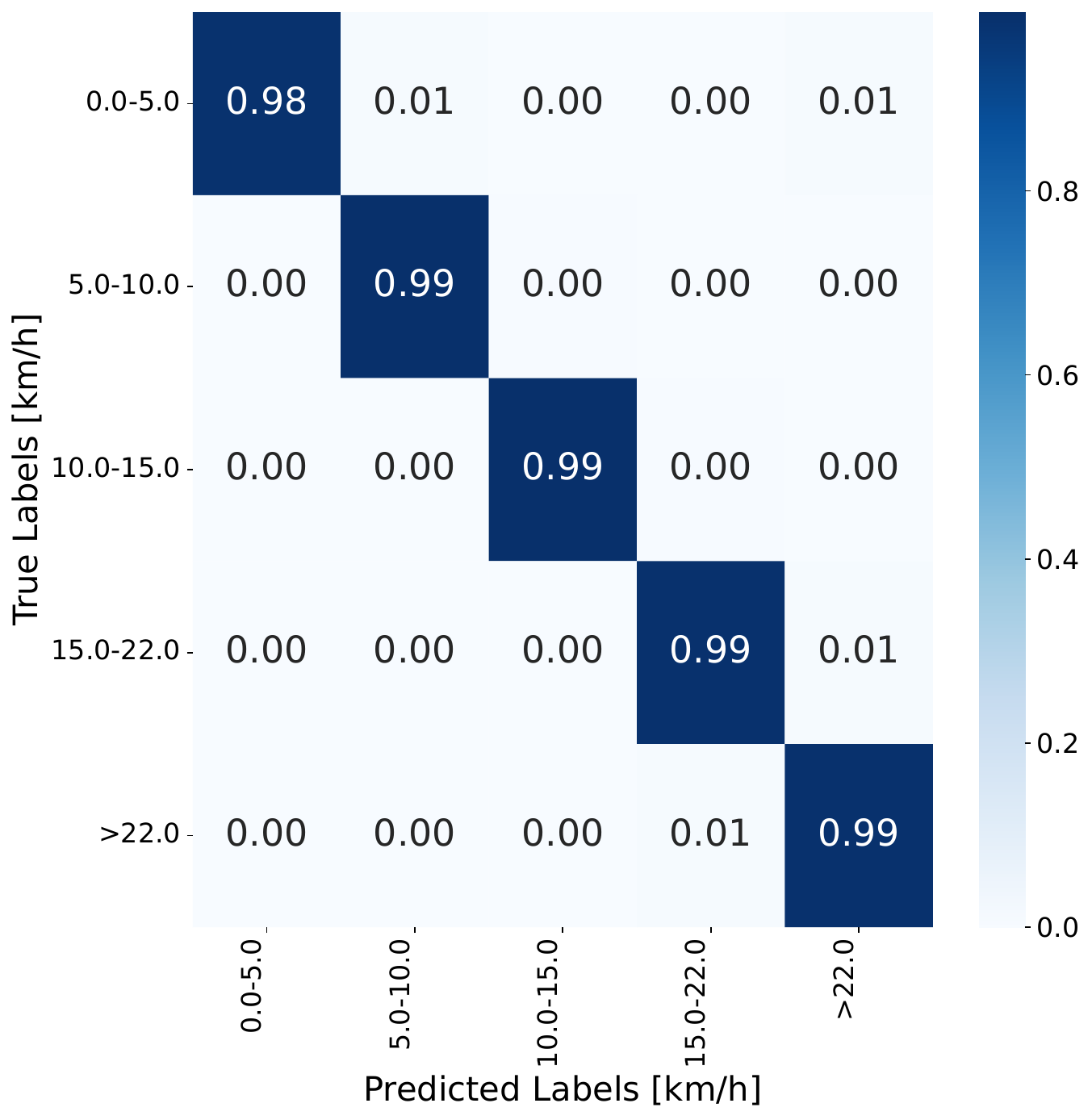}
    \caption{Wind speed classification accuracy on the test set.}
    \label{fig:wind_accuracy}
\end{figure}
\subsection{Weather Data Collection} 
\label{sec:weather_station}
We gather the training data, i.e., the \gls{tx} and \gls{rx} \gls{ofdm} frames in the following fashion: We trigger a rain/no rain check every $2$ minutes. If it is raining, we record a measurement. If it is not raining, we record no-rain measurements every $10$ minutes. A measurement is recorded for $1$ second, gathering a total of $100$ \gls{ofdm} frames, both for \gls{tx} and \gls{rx}. The choice for this criteria is storage issues and the predominance of no-rain scenarios.

We gather the reference values for the supervised learning tasks (see \fref{sec:classification} and \fref{sec:regression}) with the use of a weather station (see \fref{fig:weather_station}) \cite{weatherstation}. The device is installed in close vicinity to the FR2 \gls{isac} \gls{poc} and periodically records measurements for precipitation rate (mm/h), wind speed (km/h), among other metrics. We pair the training data with the reference values based on closeness in time. For the classification problem (see \fref{sec:classification}), we map the acquired data to classes presented in the axes of \fref{fig:rain_accuracy} and \fref{fig:wind_accuracy}. While the aim was to balance the classes' distribution, weights are employed to ensure no class dominates on training.
\subsection{Training Parameters}
The dimensions of the \gls{csi} matrix after zero-padding, $\bar{\bH}$, are $N_{\text{pad}} = M_{\text{pad}} = 2048$. For periodogram evaluation, we crop the periodogram to a range of $\PR{0,450}$ meters and a Doppler shift of $\PR{-10,+10}$ m/s, corresponding to $N'=746$ range bins and  $M' = 68$ speed bins. For training and testing, we use $S_{\text{nrain}} = 9786$ and $S_{\text{rain}} = 9780$ samples of no-rain and rain scenarios, respectively, giving a total of $S=19566$ samples. On each of these sets, we separate $80\%$ and $20\%$ of the dataset for training and testing, respectively. For training and testing, we use $158$ and $39$ batches, respectively, with a batch size of $100$ samples. To ensure a balanced representation across scenarios, each batch is constructed with $50$ frames from the no-rain dataset and $50$ frames from the rain dataset. For training, we use the Adam optimizer. The proposed model was implemented in PyTorch (v2.5.1) with CUDA support (v12.1) and trained on an NVIDIA GPU. All results and performance metrics are computed using the test set.

\subsection{Performance Metrics}
Here we describe the performance metrics used for classification and regression, respectively.
\subsubsection{Classification}
For classification, we present our results as normalized confusion matrices, which indicate how many samples out of all samples were correctly classified in each class for each weather metric. 
\subsubsection{Regression}
For regression, we present our results as \glspl{cdf} of the estimation error $\PM{\hat{y}_{s,w}-y_{s,w}}$ for each weather metric and summarize the curves by looking at the mean and 90-percentile, which we define as $\text{Pr}\PR{Y \leq y} = 0.9$. For comparison with related work, we also look at the \gls{rmse} \cite{Diba2021}.
\subsection{Performance Results}
\begin{figure*}[t]
\centering
\setlength{\tabcolsep}{0pt} 
\renewcommand{\arraystretch}{0} 
\begin{subfigure}[b]{0.243\linewidth}
    \centering
    \includegraphics[height=4.1cm]{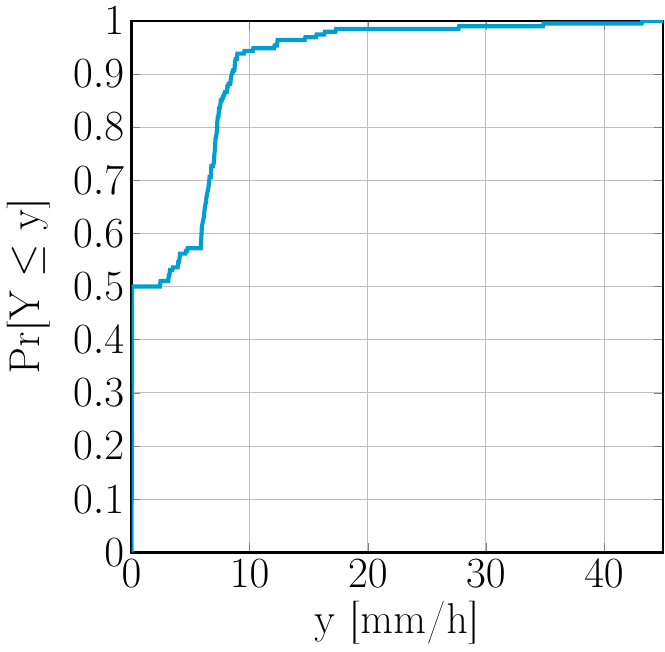}
    \caption{Precipitation rate distribution}
    \label{fig:rain_distribution}
\end{subfigure}
\begin{subfigure}[b]{0.25\linewidth}
    \centering
    \includegraphics[height=4.1cm]{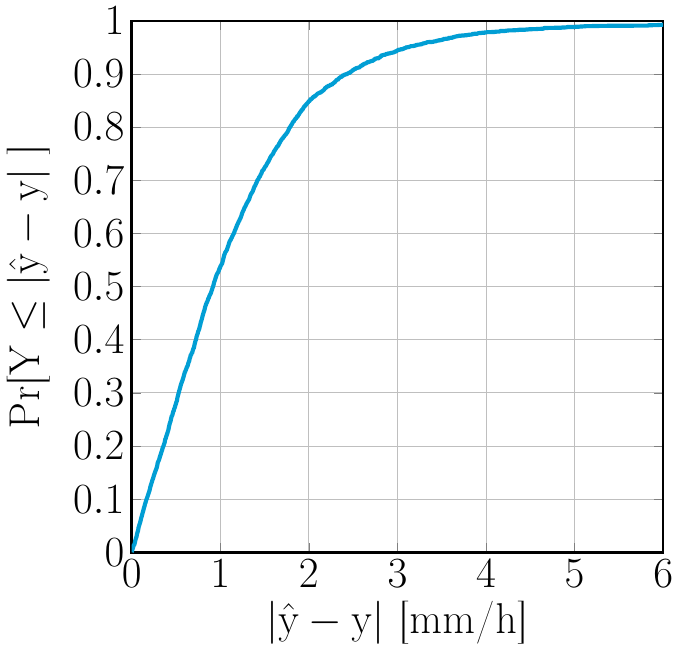}
    \caption{Precipitation\hspace{1.5pt}rate\hspace{1.5pt}estimation\hspace{1.5pt}error}
    \label{fig:rain_error}
\end{subfigure}
\begin{subfigure}[b]{0.243\linewidth}
    \centering
    \includegraphics[height=4.1cm]{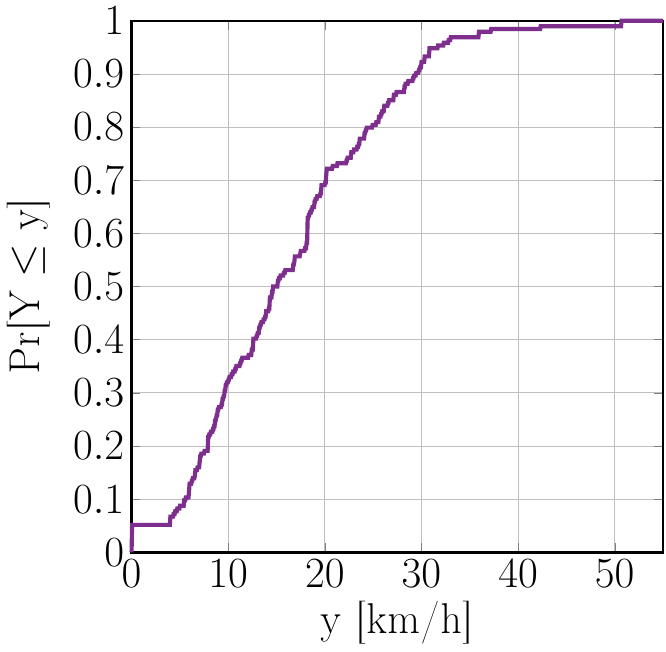}
    \caption{Wind speed distribution}
    \label{fig:wind_distribution}
\end{subfigure}
\begin{subfigure}[b]{0.243\linewidth}
    \centering
    \includegraphics[height=4.1cm]{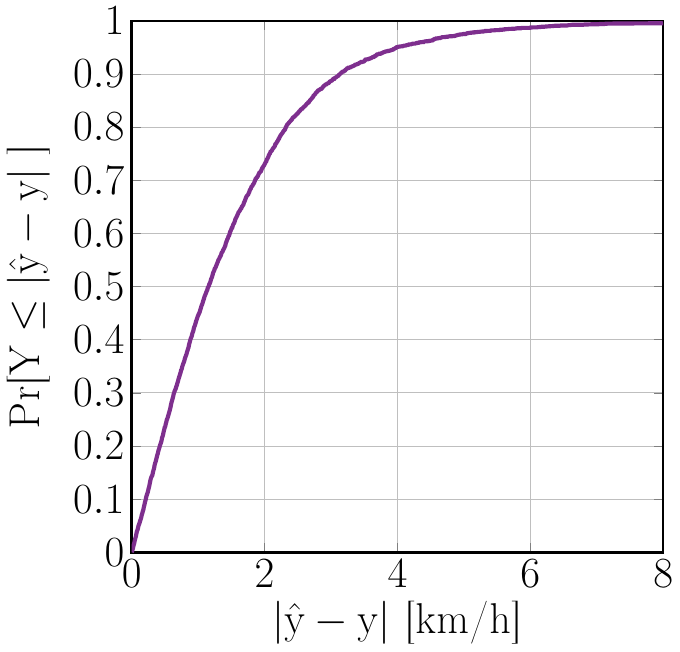}
    \caption{Wind speed estimation error}
    \label{fig:wind_error}
\end{subfigure}
\caption{\glspl{cdf} for the test set of the (a) precipitation rate distribution, (b) precipitation rate estimation error, (c) wind speed distribution, and (d) wind speed estimation error.}
\label{fig:results_regression}
\end{figure*}
\subsubsection{Classification} 
For the following experiments, we consider that $50\%$ of the samples have a precipitation rate equal to $0$ mm/h, while the other $50\%$ are distributed among varying precipitation rates.

In Figs.~\ref{fig:rain_accuracy} and \ref{fig:wind_accuracy}, we present the normalized confusion matrices for precipitation rate and wind speed, respectively, based on test set classification. For precipitation rate and wind speed, the classification accuracy is $99.38\%$ and $98.99\%$, respectively. 
\subsubsection{Regression}
In Figs.~\ref{fig:rain_distribution} and \ref{fig:wind_distribution}, we present both the precipitation rate and wind speed data distribution on the test set. As we can observe, precipitation rate {lies} within the interval of $\PR{0,43.2}$ mm/h while wind speed {lies} within the interval of $\PR{0,50.7}$ km/h. 

In Figs.~\ref{fig:rain_error} and \ref{fig:wind_error}, we present the \glspl{cdf} for both the precipitation rate and wind speed estimation errors based on inference on the test set. In our experiments, the precipitation rate achieves a \gls{mae} of $1.2$ mm/h, 90-percentile error of $2.43$ mm/h, and an \gls{rmse} of $1.76$ mm/h while the wind speed achieves an \gls{mae} of $1.5$ km/h, a 90-percentile error of $3.14$ km/h, and an \gls{rmse} of $2.04$ km/h. Compared to \cite{Diba2021} at $38$ GHz, our method achieves a 3.64 mm/h improvement in \gls{rmse} for precipitation rate estimation and additionally, provides wind speed estimation. Given the strong performance of regression, we recommend its deployment over classification in a product implementation.

%% file: sections/5-conclusions.tex
% !TEX root = weather_sensing_paper.tex
% DO NOT REMOVE THE ABOVE COMMENT!

\section{Conclusions and Future Work}
\label{sec:conclusions}
We have proposed a solution to enable weather sensing in \gls{6g} \gls{isac} networks, and have validated our approach with extensive measurements on a realistic setup. We have performed weather estimation using \gls{csi}-based periodograms as inputs to two supervised learning tasks, classification and regression, implemented as a \gls{cnn}. The proposed method has enabled us to jointly detect and estimate weather phenomena such as precipitation and wind, solely using communication signals.

Our experiments with measurements from the FR2 \gls{isac} \gls{poc} have demonstrated that we are capable of achieving accuracies of $99.38\%$ and $98.99\%$ for the classification of precipitation rate and wind speed, respectively. We have also shown that by using our method, we can achieve an \gls{mae} of $1.2$ mm/h and $1.5$ km/h for precipitation rate and wind speed, respectively. These results show good promise for the deployment of weather sensing in future \gls{6g} \gls{isac} networks.

There are several avenues for future work. First, improving the generalization of our method through transfer learning and integration of measurements from other sites. Second, performing experiments in other frequency bands, e.g., FR1 ($f_c = 3.5$ GHz). Third, considering a fully mono-static configuration, where the \gls{ru} operates in full-duplex mode. Fourth, performing a quantitative study to evaluate the system's parameters.